\documentstyle[twoside,fleqn,espcrc2]{article}

\def\apj{Ap.~J.}
\def\apjl{Ap.~J.~(Letters)}

\def\sles{\lower2pt\hbox{$\buildrel {\scriptstyle <}
   \over {\scriptstyle\sim}$}}
\def\sgreat{\lower2pt\hbox{$\buildrel {\scriptstyle >}
   \over {\scriptstyle\sim}$}}
\def\undertext#1{$\underline{\smash{\hbox{#1}}}$}

\def\msol{\,M_{\odot}}
\def\sgreat{\lower2pt\hbox{$\buildrel {\scriptstyle >}
   \over {\scriptstyle\sim}$}}

\def\mj{$\,{\rm M}_{\rm J}\,$}
\def\mjj{{\rm M}_{\rm J}\,}

\def\etal{{\it et~al.}}

\def\lo{$L_\odot\,$}
\def\mo{$M_\odot\,$}
\def\mp{$\,M_p$}

\def\lbol{$L_{bol}\,$}
\def\mstar{$M_{\ast}$}
\def\Dwa{$\,$\uppercase\expandafter{\romannumeral5}$\,$}

\def\wig#1{\mathrel{\hbox{\hbox to 0pt{%
          \lower.5ex\hbox{$\sim$}\hss}\raise.4ex\hbox{$#1$}}}}

\def\rj{R$_{\rm J}\ $}
\def\arcsec{^{\prime\prime}}
\def\arcsecb{^{\prime\prime}}
\title{Theoretical Models of Extrasolar Giant Planets\thanks{The authors would like to acknowledge the
                support of NASA via grant NAGW--2817 and of the NSF via grant
                AST93-18970.}}

\author{A. Burrows\address{Department of Astronomy, University of Arizona, 
        Tucson, AZ 85721}
        W.B. Hubbard\address{Department of Planetary Sciences, University of Arizona, 
        Tucson, AZ 85721}
        J.I. Lunine$^{\rm b}$ 
        T. Guillot$^{\rm b}$
        D. Saumon$^{\rm b}$
        M. Marley\address{Department of Astronomy, New Mexico State University,
                  Box 30001/Dept. 4500, Las Cruces NM 88003}
        and
        R.S. Freedman\address{Sterling Software, 
                  NASA Ames Research Center, Moffett Field CA 94035}
} 

\begin{document}

\begin{abstract}
   The recent 
   discoveries
   of giant planets around nearby stars has galvanized the planetary science community,
   astronomers, and the public at large. 
   Since {\it direct} detection is now feasible, and is
   suggested by the recent acquisition of Gl229 B, it is crucial for the 
   future of extrasolar planet
   searches that the fluxes, evolution, and physical structure of objects 
   from Saturn's mass to 15 Juipter masses
   be theoretically investigated.
   We discuss our first attempts to explore
   the characteristics of extrasolar giant planets (EGPs),
   in aid of both NASA's and ESA's recent plans to search for such planets around nearby stars.
\end{abstract}

\maketitle

\section{Introduction}

After years of obscurity in the backwaters of astronomy, the recent epochal detections
of giant planets around nearby stars has put the search for extrasolar
planets on its center stage.  The discoveries of 51 Peg B \cite{mq95}, $\tau$ Boo B, 55 Cnc B and C,
70 Vir B, and 47 UMa B \cite{mb96a,mb96b,bm96}, 
and the belated recognition that HD114762 \cite{lat89}  
is in this class have galvanized the planetary science community, astronomers, and the public at large.
Table 1 lists these newly--discovered planets, in order of increasing semi--major axis, along
with the giant planets in our solar system and the brown dwarf Gl229 B \cite{nak95}.
Also shown are $M_p sin (i)$, orbit period, eccentricity, distance to the sun, estimated surface temperature, and
age (when an age could be comfortably assigned).  The wide range in mass (0.5\mj to $\sim$ 10\mj)
and period (3.3 {\it days} to $\sim$ 20 years), as well as the proximity of many of the planets
to their primaries, was not anticipated by most planetary scientists.
Though the technique of Doppler spectroscopy used to find most of these planets 
selects for massive, nearby
objects, their variety and existence is a challenge to conventional
theory.  

Within the last year, building upon our previous experience in the modeling of brown dwarfs and M stars,
we published theoretical studies on the evolution and spectra of extrasolar giant planets (EGPs\footnote[1]{We use this
shorthand for \undertext{E}xtrasolar \undertext{G}iant \undertext{P}lanet, but the terms ``exoplanet'' or ``super--jupiter''
are equally good.})
(Burrows \etal\ \cite{burr95}; Saumon \etal\ \cite{saumon96}; Guillot \etal\ \cite{guillot96}; Marley \etal\ \cite{marley96}).
Brown dwarfs ($< 0.08 \msol$) are transition objects
that straddle the realms of planets and stars, have atmospheres
composed of a complex soup of molecules and grains, and can achieve central
temperatures adequate to consume their stores of cosmological deuterium, while
inadequate to ignite sufficient light hydrogen to avoid cooling into
obscurity within a Hubble time.
Nevertheless, they are characterized by straightforward
extensions of the same equation-of-state (EOS) and atmospheric physics
appropriate for the less massive giant planets.

\begin{table*}[hbt]
\setlength{\tabcolsep}{0.65pc}
\newlength{\digitwidth} \settowidth{\digitwidth}{\rm 0}
\caption{The Giant Planet Bestiary}
\begin{tabular*}{\textwidth}{@{}@{\extracolsep{\fill}}lllllllll}
\hline
Object & Star & Mass (M$_J$) & $a$ (A.U.) & P (days) & $e$ & T$_{\rm eff}$ (K) & D (pc) & Age (Gyrs) \\
\hline
$\tau$ Boo B & F7 & \sgreat\ 3.87 & 0.046 & 3.313 & 0.0 & 1500 & 19 & ? \\
51 Peg B & G2.5 & \sgreat\ 0.46 & 0.05 & 4.23 & 0.0 & 1250 & 15.4 & 8 \\
$\upsilon$ And B & F8 & \sgreat\ 0.6 & 0.057 & 4.61 & $\sim$0.03 & 1350 & 16.1 & ? \\
55 Cnc B&G8 & \sgreat\ 0.8 & 0.1  & 14.76 & 0.0 & 1000 & 13.5 & ? \\
HD114762 & F8 & \sgreat\ 9 & 0.38 & 84 & 0.33 & 450 & 28 & ? \\
70 Vir B & G4 & \sgreat\ 6.6 & 0.45 & 116.6 & 0.40 & 380 & 9 & 8 \\
47 UMa B & G0 & \sgreat\ 2.39 & 2.11 & 3 yrs & $\sim$0.03 & 180 & 12 & 6.9 \\
Gl411 B (?) & M2 & $\sim$0.9 & 2.33 & 5.8 yrs & ? & 100 & 2.52 & ? \\
55 Cnc C & G8 & \sgreat\ 5 & 5--10 & 15--20 yrs & ? & 175 & 13.5 & ? \\
Jupiter & G2 & 1.0 & 5.2 & 11.86 yrs & 0.048 & 125 & -- & 4.6 \\
Saturn & G2 & 0.3 & 9.5 & 29.46 yrs & 0.056 & 95 & -- & 4.6 \\
Gl229 B & M1 & 30--55 & \sgreat\ 44.0 & \sgreat\ 400 yrs & ? & 960 & 5.7 & \sgreat\ 1 \\
\hline

\end{tabular*}
\end{table*}

Some of the space platforms and new ground--based facilities that have or will
obtain relevant infrared and optical data include
the HST (WFPC2, NICMOS), the IRTF, the MMT 6.5-meter upgrade, the Large Binocular Telescope (LBT)
(planned for Mt. Graham), Keck's I and II (with HIRES), the European ISO, UKIRT,
and SIRTF, along with a large number of medium-- to large--size telescopes
optimized or employed in the near--infrared.  
One project of Keck I and II,
under the aegis of NASA's ASEPS-0 (Astronomical Study of Extrasolar Planetary Systems)
program, will be to search for giant planets around nearby stars.
A major motivation for the Palomar Testbed Interferometer (PTI) supported by NASA
is the search for extrasolar planets.
Recently, Dan Goldin, the NASA administrator, outlined a program to detect planetary
systems around nearby stars that may become a future focus of NASA.
This vision is laid out in the {\it Exploration of Neighboring Planetary Systems (ExNPS)
Roadmap} (see also the ``TOPS'' Report \cite{tops92}).

\begin{figure*}
\vspace*{2.5in}
\hbox to\hsize{\hfill\includegraphics{egpfig1.ps}\kern+0in\hfill}
\caption{Bolometric luminosity
(\lbol) in solar units of a suite of EGPs placed at a
distance of 5.2 A.U. from a G2\Dwa star
versus  time ($t$) in Gyr.  The reflected luminosity is not included, but
the absorbed component is. At $t \sim 0.2\,$Gyr, the luminosity
of the 14\mj EGP exceeds that of the 15\mj EGP because of
late deuterium ignition.
The data point at $4.55\,$Gyr shows the
observed luminosity of Jupiter.  The 0.3\mj EGP
exhibits a strong effect of warming by the G2\Dwa primary star at
late stages in its evolution.
Although this model resembles
Saturn in mass, here it is placed at the distance of Jupiter from
its primary.
(The flattening in $L$ vs. $t$ for low masses and great ages is a
consequence of stellar insolation.)
The insert shows, on an expanded scale, the comparison
of our lowest-mass evolutionary trajectories with the present
Jupiter luminosity (from \cite{burr95} ). }
\end{figure*}

\section{Early Calculations of the Evolution and Structure of Extrasolar Giant Planets}

Our group \cite{burr95,saumon96} 
recently calculated 
a suite of models of the evolution and emissions
of EGPs.
Surprisingly, 
no one had
accurately mapped out the properties of
objects between the mass of giant planets in our solar system and the traditional
brown dwarfs ($>10 - 20$ $\mjj$, where $\mjj$ is the mass of
Jupiter, $\sim$ 10$^{-3} \msol$).  This is precisely the mass range for the newly--discovered planets listed in Table 1.

EGPs will radiate in the optical by reflection and in the
infrared by the thermal emission of both absorbed stellar light and the planet's
own internal energy.
In Burrows \etal\ \cite{burr95} we included the effects of ``insolation'' by a central star of mass \mstar$\,$ and considered
semi-major axes ($a$) between 2.5 A.U. and 20 A.U.  Giant
planets may form preferentially near 5 A.U. \cite{boss95}, but as the new data dramatically affirm,  
a broad range of $a$s can not be excluded.
We evolved EGPs with masses ($M_p$) from 0.3\mj $\,$(the mass of Saturn)
through 15\mj .
Whether a 15\mj object is a planet or a brown dwarf is largely a semantic issue, though one might
distinguish gas giants and brown dwarfs by their
mode of formation (e.g. in a disk or ``directly'').  Physically, compact hydrogen-rich objects with masses from 0.00025 \mo$\,$ through
0.25 \mo$\,$ form a continuum.  However, our EGPs
above $\sim$13 \mj$\,$ do burn ``primordial'' deuterium for up to 10$^{8}$ years. 

The evolution of the bolometric luminosity (\lbol) of our suite of EGPs orbiting 5.2 A.U. from a G2\Dwa star
as calculated in Burrows \etal\ \cite{burr95} is depicted in Figure 1.  One is
struck immediately by the high \lbol 's for early ages and high masses.
That a young ``Jupiter'' or ``Saturn'' will be bright has
been known for some time, but ours are the first detailed calculations for
objects with \mp $\ >$ \mj and ages, $t$, greater than 10$^{7}$ years.
Below about 10\mj, \lbol$\,$ is very roughly proportional
to $M_p^{\alpha}\!/t^{\beta}$, where $1.6 \le \alpha \le 2.1$  and $1.0 \le \beta \le 1.3$. An EGP with a mass of 2 \mj at age
10$^{7}$ years is two thousand times brighter than the current Jupiter (and its $T_{\rm eff}$ is $\sim$700 K).
At the age of the Pleiades ($\sim7\times10^{7}$ years), such an EGP would be $\sim 200$ times brighter
(with $T_{\rm eff}$ $\sim 420\,$K) and at the age of the Hyades ($\sim6\times10^{8}$ years) it would be $\sim 18$ times
brighter (with $T_{\rm eff}$ $\sim 235\,$K).  The measured  \lbol and $T_{\rm eff}$ of Jupiter are
$2.186\pm0.022\times10^{-9}$ \lo and  $124.4\pm0.3\,$K, respectively \cite{pearl91}.  At an age of $4.55\,$Gyr, our
model of Jupiter has a luminosity of $2.35\times10^{-9}$ \lo and an effective temperature of $122\,$K.

A few ``facts'' will serve to illustrate the mid-infrared character of massive young EGPs.
Since the fluxes shortward
of $10\, \mu $m (the N band) are generally on or near the Wien tail of the EGP spectrum, the
fluxes in the near- and mid-infrared spectral bands increase even faster
with mass and youth than \lbol .  In particular, if we assume that the emission is {\it Planckian}, that the orbital separation
is 5.2 A.U., and
that \mstar$\,$ equals 1.0$\,M_\odot$, Jupiter's N band flux would be $\sim 8000$ times higher at age
$10^{7}$ years than it is now.  At the age of our solar system,  a 2\mj EGP and a 5\mj EGP would be $\sim6$  and $\sim90$
times  brighter in the N band than the current Jupiter.
Furthermore,
in the M band ($\sim5\,\mu $m), a 2\mj\ EGP would be $\sim\!60,\!000$ times brighter at $10^{7}$
years than the current Jupiter, but at $10^{9}$ years ``only'' $\sim2.5$ times brighter than a coeval Jupiter.  At the
age of the Hyades, Saturn would be as bright as the current Jupiter.  The fluxes at Earth
due to the thermal emissions shortward of $10\,\mu $m of EGPs in the Pleiades ($D\sim125\,$ parsecs)
would be greater than those from EGPs in the Hyades ($D\sim45$ parsecs), despite the latter's
relative proximity, because the Pleiads are younger (and, hence, at higher $T_{\rm eff}$).

\section{Giant Planets at Small Orbital Distances}

It had been thought that the orbital radius of a giant planet had to be at least 4.0 A.U. \cite{boss95}, but the epochal detections of
$\tau$ Boo B, 55 Cnc B, and 51 Peg B belie this paradigm.
Amazingly,  51 Peg B \cite{mq95}
is orbiting a G2.5 star, 
with a 4.23--{\it day} period, a semi-major axis of 0.05 A.U., an eccentricity near zero,
and an inferred mass between 0.5 and 2 Jupiter masses. 

One hundred times closer to its primary than Jupiter itself, 51 Peg B thwarts conventional wisdom.
Boss \cite{boss95} had argued that the nucleation of a H/He-rich Jovian planet around a rock and ice core
could be achieved in a protostellar disk only at and beyond the ice point (at $\sim$160 K) exterior
to 4 A.U.~\  Walker \etal\ \cite{walk95} had surveyed 21 G-type stars for reflex motion over 12 years,
had detected none, and had derived upper limits
of 0.5--3 \mj for the masses (modulo sin(i)) of any planets interior to $\sim$6 A.U.~\ that they may have missed.
Zuckermann, Forveille, \& Kastner \cite{zuck95}
had measured CO emissions from a variety of near--T Tauri disks, had extrapolated to H$_{2}$, and had
concluded that there may not be enough mass or time to form a Jupiter around a majority of stars.
The discoveries of 51 Peg B, $\tau$ Boo B, and 55 Cnc B, while not strictly inconsistent with any of these papers, vastly enlarge
the parameter space within which we must now search.

\begin{figure*}
\vspace*{2.5in}
\hbox to\hsize{\hfill\includegraphics{pegfig2.ps}\kern+0in\hfill}
\caption{Hertzprung-Russell diagram
for 1$\,$M$_{\rm J}$ planets orbiting at
0.02, 0.025, 0.032, 0.05, and 0.1$\,$A.U. from a star with the properties of 51 Peg A,
assuming a Bond albedo of 0.35.
Arrows indicate the corresponding equilibrium
effective temperature. A Jupiter model is also shown, the diamond
in the bottom right-hand corner corresponding to the present-day effective
temperature and luminosity
of the planet. Evolutionary tracks for planets of solar composition are
indicated by lines connecting dots
which are equally spaced in log(time). The numbers 7, 8, 9, 10 are the common logarithms of the
planet's age.
Zero-temperature models for 1$\,$M$_{\rm J}$ planets made of olivine
(Mg$_2$SiO$_4$) are indicated by triangles.
The Hayashi forbidden region,
which is enclosed by the evolutionary track of the fully convective model, is
shown in dark grey. Models in the light grey region have
radii above the
Roche limit (and therefore are tidally disrupted by the star).
The region where classical Jeans escape becomes significant
is bounded by the dash-dotted line. Lines of constant radius are
indicated by dotted curves. These correspond, from bottom to top, to
radii (in units of \rj) in multiples of 2, starting at 1/4
(from \cite{guillot96} ).
}
\end{figure*}

After 8
billion years (the estimated age of 51 Peg A), if 51 Peg B is a gas giant, its radius is only 1.2$\,$R$_{\rm J}$ 
(where \rj is the radius of Jupiter) and its
luminosity is about $3.5\times 10^{-5}\,L_\odot$.  This bolometric luminosity is more than $1.5 \times 10^4$
times the present luminosity of Jupiter and only a factor of two below that at the edge of the main sequence.
A radiative region encompasses
the outer 0.03\% in mass, and 3.5\% in radius. 
Figure 2 from Guillot \etal\ \cite{guillot96} is a theoretical Hertzprung-Russell 
diagram that portrays the major results
of our 51 Peg B study.  Depicted are
$L$--$T_{\rm eff}$ tracks
for the evolution of a 1 \mj gas giant and $L$
for a 1 \mj  ``olivine'' planet, all at a variety of orbital distances (indicated by the arrows).
Also shown are the Hayashi track (boundary of the
dark shaded region), the Hayashi
exclusion zone (the dark shaded region itself), the Roche exclusion zone (the lightly shaded region),
and the classical Jeans evaporation limit (dash-dotted line).  
The dotted lines on Figure 2 are lines of constant radius.
The numbers on the tracks are the common logarithms of the ages in years.
The study by Guillot \etal\ \cite{guillot96} demonstrated that 51 Peg B is well within its Roche lobe
and is not experiencing significant photoevaporation.  Its deep potential well ensures that,
even so close to its parent, 51 Peg B is {\bf stable}.
If 51 Peg B were formed beyond an A.U. and moved inward on a timescale greater than $\sim 10^{8}$ years, it would
closely follow the $R_{\rm p} \sim$ \rj trajectory to its equilibrium position.

For radiative/convective gas-giant models of 51 Peg B, the predicted radii after
1 Gigayear (Gyr) are between 1.35 \rj and 1.9 \rj for $M_{\rm p}$'s from 2.0 \mj$\,$ to 0.5 \mj $\,$.
These are as much
as a factor of two smaller that the corresponding radii for fully convective planets. After 8 Gyr, the radii
for these same planets are between 1.2 \rj and 1.4 \rj.
A giant terrestrial planet  with a mass
between 0.5 \mj and 2.0 \mj would have a radius between 0.31 \rj and 0.35 R$_{\rm J}$,
three times smaller than that of a gas giant in the same mass range, and its corresponding 
luminosity would be an order of magnitude
lower ($2.0-2.5\times 10^{-6}\,L_\odot$).  
If photometry can be performed on 51 Peg B, $\tau$ Boo B, or 55 Cnc B, a measurement of bolometric luminosity will
immediately distinguish the different models.

\section{Atmospheric, Evolutionary, and Spectral Models of Gl229 B}

To constrain the properties of the brown dwarf Gl229 B \cite{nak95,opp95,geb96,matt96},
we (Marley \etal\ \cite{marley96})
recently constructed a grid of brown dwarf model
atmospheres with $T_{\rm eff}$ ranging from 600 to 1200 K and
$100<g<3200$ m s$^{-2}$.  
We assumed a standard solar composition for the bulk of
the atmosphere.
Refractory elements (for example Fe, Ti, and silicates) condense deep in the
atmosphere for $T_{\rm eff} \approx 1000$ K, and thus
have negligible gas-phase abundance near the photosphere, as
is also true in the atmosphere of Jupiter.
For an atmosphere similar to that of Gl229 B, chemical equilibrium
calculations indicate that C, N, O, S, and P
are found mainly in the form of methane (CH$_4$), ammonia (NH$_3$),
water (H$_2$O), hydrogen sulfide (H$_2$S), and phosphine (PH$_3$),
respectively.   However, deep in the atmosphere, chemical equilibrium
favors CO over CH$_4$ and $\rm N_2$ over $\rm NH_3$.
Our model atmospheres incorporate
opacities of these molecules, H$_2$, and He in their respective
solar abundances and includes no other elements.

\begin{figure*}
\vspace*{2.85in}
\hbox to\hsize{\hfill\includegraphics{billshade.ps}\kern+0in\hfill}
\caption{The grey shaded area 
shows the best--fit region for Gl229 B.
Solid lines depict the evolution of $T_{\rm eff}$ and $g$
as various mass brown dwarfs cool.  The masses in jupiter 
masses are indicated near the appropriate lines.  Several contours of
constant radius (long-dashed curves) and constant age
(short-dashed curves) are also shown (from \cite{marley96} ).
}
\end{figure*}

In Marley \etal\ \cite{marley96}, we employed a stellar evolution code and atmosphere models to
estimate the physical properties of the brown dwarf, Gl229 B.
By comparing our theoretical spectra with the UKIRT \cite{geb96} and 
HST \cite{matt96} data, we derived an effective temperature of
$960 \pm 70$ K and a gravity between $0.8 \times 10^{5}$ and $2.2 \times 10^{5}$ cm s$^{-2}$.
These results translate into masses and ages of 30--55 \mj and 1--5 Gyr, respectively.
As Figure 3 indicates, gravity maps almost directly
into mass, and ambiguity in the former results in uncertainty in the latter.
While the near infrared spectrum of Gl229 B is dominated by $\rm H_2O$,
we confirmed the presence of $\rm CH_4$ in the atmosphere from our modeling of its features at
1.6--1.8 $\mu$m, 2.2--2.4 $\mu$m, and 3.2--3.6 $\mu$m.
In addition, we found a flux enhancement in
the window at 4--5 $\mu$m throughout the T$_{\rm eff}$ range from 124 K (Jupiter)
through 1300 K, and, hence, that this band is a {\bf universal} diagnostic for brown dwarfs
and planets. By comparison,
the widely-used K band at 2.2 $\mu$m is greatly suppressed by strong CH$_4$ and
H$_2-$H$_2$ absorption features.
Beyond 13 $\mu$m, the decreasing flux falls slightly more rapidly than a
Planck distribution with a brightness temperature near 600 K.

\begin{figure*}
\vspace*{2.7in}
\hbox to\hsize{\hfill\includegraphics{billfig5.ps}\kern+0in\hfill}
\caption{Spectral dependence of the
flux received at the Earth  from
extra-solar giant planets (EGPs) orbiting at 5.2 A.U. from a G2\Dwa
star at 10 parsecs from the Earth (The orbit subtends an angle
of $\sim$0.5$^{\prime\prime}$), according to \cite{burr95}.
Objects of 1 \mj (solid curves)
and 5 \mj (dashed curves) are displayed at the
following ages: log $t ({\rm yr})=$ 7.0, 7.5, 8.0, 8.5, 9.0,
9.5, 9.7 (from left to right).  The reflected component of the
light is essentially independent of the mass and the age of the EGP.
The EGP is assumed to emit like a {\it blackbody} and to reflect incident
light as a grey body.  Standard photometric bandpasses are shown at
the top.  Also shown are the design sensitivities of several
astronomical systems for the detection of point sources with
a signal-to-noise ratio of 5 in a 1-hour integration (40 minutes
for NICMOS). These systems are the LBT and
MMT (solid circles and square, respectively), the three cameras
of NICMOS (open triangles, 3-pointed stars and solid triangles),
SIRTF (solid bars), and Gemini and SOFIA (dashed bars).
The spectrum of a G2\Dwa star was provided by A. Eibl (private
communication, 1995).
It should be stressed that while SIRTF is unlikely to have the {\it
angular resolution} to detect the EGPs of this example, the same EGPs
at slightly larger separations around a star that is slightly closer
will be well within its detection envelope.
}
\end{figure*}

\section{Potential for Direct Detection of EGPs}

Imaging giant planets around nearby stars presents major technological challenges.  The
difficulties arise mainly from the following problems:
1) The brightness ratio
between the star and the planet is large and ranges from $\sim 30$ to 10$^9$,\  2) The
angular separation
may range from $5\arcsec$ in the favorable case to a daunting $0.002\arcsec$, and \  3) The
flux from the planet is generally very low.  These three factors stretch the current limits of optical and
infrared technologies.  Current and next-generation instruments
are reaching sensitivity levels within the range of our predicted fluxes of
EGP's, but they do not always have sufficiently high angular resolution to resolve
the EGP from its parent star.  (The separations for the newly--discovered planets listed in Table 1 range
from $0.002\arcsec$ for $\tau$ Boo B to $0.5\arcsec$ for 55 Cnc C.) 
The issue of angular resolution is further complicated by the
problem of light scattered in the telescope optics.  The point-spread function of
diffraction-limited optical systems typically has a very faint halo which can spread over
several arcseconds around the Airy disk, due to minute residual errors in the
figure of the mirror, light scattered inside the telescope,
or residual atmospheric distortions of the images.  Because of the enormous contrast
between the planet and the primary star, the signal of the planet can be  lost
in the halo of the primary star.  The brightness of this faint halo
is very difficult to predict and is expected to vary widely from one instrument to another.

Detection of planets by direct techniques, {\it i.e.} imaging using adaptive optics
or interferometric techniques, must also take into account the scattering
of light by dust systems, analogous to our zodiacal light, around
candidate stars. Further, such imaging in the mid-infrared is inhibited
by our own zodiacal dust, requiring that infrared interferometers be
placed in heliocentric orbits at 3$\,$A.U. or beyond to avoid the worst of the
dust emission (ExNPS report and R. Angel, personal communication).

Ignoring for the moment the question of angular resolution, Figure 4 compares the
sensitivities of various ground--based and space--based telescopes with some theoretical
fluxes calculated in \cite{burr95} and \cite{saumon96} under the blackbody assumption. 
These comparisons help us gauge the capabilities of various platforms
for giant planet searches and has led us to conclude that current technology
will indeed be up to the challenge.
Figure 4 depicts the flux in Janskys versus wavelength at 10 parsecs
for a 1\mj and a 5\mj$\,$ EGP that are 5.2 A.U. from a G2\Dwa
star (a solar analog), at times between $10^7$ and $5\times10^{9}$ years.  
The reflected component is included in Figure 4.
Also shown on Figure 4 are the $5\sigma$ point-source sensitivities at various wavelengths for
the Space InfraRed Telescope Facility (SIRTF, \cite{ew92}), 
the Large Binocular Telescope (LBT, \cite{a94}), 
the upgraded ``Multiple Mirror'' Telescope (MMT, \cite{a94}), Gemini \cite{mount94},
the Stratospheric Observatory For Infrared Astronomy (SOFIA, \cite{e92}), and NICMOS \cite{thomp92}.

The three NICMOS (Near Infrared Camera and Multiple Object Spectrograph) cameras,
with resolutions of 0.043, 0.075 and 0.2$\arcsecb\,$
respectively, are very promising instruments for the detection of EGP's in the solar
neighborhood.  They are sensitive in the near infrared where most
EGP's emit in reflected light.
With the adaptive optics scheme proposed by Angel \cite{a94},
both the MMT and the LBT will achieve diffraction-limited resolution ($\sim 0.025\arcsecb$ and
$\sim 0.014\arcsecb$, respectively, at 0.8$\,\mu$m) from the ground.  Measurable star/planet flux
ratios might be as high as $\sim 10^9$ for bright enough stars.
Hence, the two telescopes will have sensitivities comparable to the NICMOS cameras at $\lambda
=0.8\,\mu$m and may successfully tackle the problem of scattered light.
Gemini and SOFIA 
are sensitive at mid- to far-infrared wavelengths.
This band spans the thermal emission of EGP's, and the signal is
predicted to be relatively insensitive to the
type of central star.  The sensitivity of SOFIA
is too low to be useful at wavelengths beyond 10$\,\mu$m.  The lower angular resolution
of these telescopes ($\sim 1\arcsecb$ at best) limits useful searches to nearby
systems with fairly large orbital radii.

SIRTF will have the highest angular resolution
of all space-based instruments in the mid- to
far-infrared.  Its high sensitivity gives it a real chance of detecting the thermal
emission of EGP's in the solar neighborhood.   SIRTF should be particularly good at searching
for EGP's around M dwarfs which are too faint in reflected light to be seen by other
powerful instruments such as NICMOS and the LBT.  Its expected angular resolution
of $\sim 1$ -- $2\arcsec$ \ limits searches to favorable combinations of
distance $D$ and orbital radius $a$.
ISO (Infrared Space Observatory) was launched in 1995 and its 5 -- 20$\,\mu$m sensitivity
is not much lower than that of SIRTF. The angular resolution of
ISO is limited by its small aperture (0.6$\,$m) and is further compromised by pointing
jitter of $2.8\arcsecb$ (ISO Observer's Manual \cite{ISO94}).

As is indicated in Figure 4, the LBT and NICMOS have the flux sensitivity to see at 10 parsecs the reflected light of
such EGPs at 5 A.U. at any age.  At the diffraction limit, these instruments
will also have the requisite angular resolution.
At 10 parsecs, SIRTF has the flux sensitivity between $5\,\mu $m and $10\,\mu $m
to detect the thermal emissions
of both a 5 \mj EGP, for ages less than $10^9$ years,
and a Jupiter at 10 A.U., for ages less
than $10^{8}$ years.

\section{The Future}

We have already preformed the first rudimentary calculations
of the atmospheres, evolution, and spectra of extrasolar giant planets, in a variety of
stellar environments.  However, much remains to be done.  Burrows \etal\ \cite{burr95}
and Saumon \etal\ \cite{saumon96} assumed that the time-- and mass--dependent fluxes from EGPs 
were Planckian.  As our recent work on Gl229 B has shown, this assumption can be
an order--of--magnitude off in the J, H, L$^{\prime}$, and M bands.  
We are planning to construct an extended grid of 
non--grey boundary conditions and will soon perform evolutionary
calculations, not only in the Gl229 B mass range (30 -- 55 \mj ), but in the 
EGP mass range from 0.3 \mj to 15 \mj (of relevance to the 
newly--discovered giant planets in Table 1).  In this way, we will derive in a self--consistent fashion the spectra and colors
of EGPs as a function of mass and age.   This expanded theory will provide a means 
of working back from a measured spectrum, T$_{\rm eff}$, and gravity to the physical characteristics of the planet itself,
assessible in no other way.  Spectra and colors are worth little if they can't be 
attached to masses, ages, and compositions.   
A complete and self--consistent theory of the evolution, emissions, and structure of extrasolar giant planets 
will be a crucial prerequisite
for any credible direct search of nearby stars. 
The present pace of giant planet discovery and NASA's and ESA's \cite{leger93} future plans
for planet searches suggest that
many more objects in the Jovian mass range (and above)
will soon be identified and subject to spectroscopic examination.


\begin{thebibliography}{99}

\bibitem{mq95} M. Mayor, M. \& D. Queloz, Nature, 378 (1995) 355.

\bibitem{mb96a}G.W. Marcy \& R.P. Butler, \apjl, 464 (1996) L147.

\bibitem{mb96b} G.W. Marcy \& R.P. Butler, public communication (1996).

\bibitem{bm96} R.P. Butler \& G.W. Marcy, \apjl, 464 (1996) L153.

\bibitem{lat89} D.W. Latham, T. Mazeh, R.P. Stefanik, M. Mayor, \& G. Burki, Nature, 339 (1989) 38.

\bibitem{nak95} T. Nakajima, {\it et al.}, Nature, 378 (1995) 463.

\bibitem{burr95} A. Burrows, D. Saumon, T. Guillot, W.B. Hubbard,  \& J.I. Lunine,
Nature, 375 (1995) 299.

\bibitem{saumon96} D. Saumon, W.B. Hubbard, A. Burrows, T. Guillot, J.I. Lunine, \& G. Chabrier, \apj, 460 (1996) 993.

\bibitem{guillot96} T. Guillot, A., Burrows, W.B. Hubbard, J.I. Lunine,  \& D. Saumon, \apjl, 459 (1996) L35.

\bibitem{marley96} M.S. Marley, D. Saumon, T., Guillot, R., Freedman, W.B. Hubbard, A. Burrows, \& J.I. Lunine,
Science, 272 (1996) 1919.

\bibitem{tops92} TOPS: Toward Other Planetary Systems, NASA Solar System Exploration
Division, Washington, D.C., 1992.

\bibitem{boss95} A.P. Boss, Science, 267 (1995) 360.

\bibitem{pearl91} J.C. Pearl \& R. A. Conrath, J. Geophys. Res. Suppl., 96 (1991) 18921.

\bibitem{walk95} G.A.H. Walker, \etal\, Icarus, 116 (1995) 359.

\bibitem{zuck95} B. Zuckermann, T. Forveille, \& J.H. Kastner, Nature, 373 (1995) 494.

\bibitem{opp95} B.R. Oppenheimer, S.R. Kulkarni, K. Matthews, \& T. Nakajima, Science, 270 (1995) 1478.

\bibitem{geb96} T.R. Geballe, S.R. Kulkarni, C.E. Woodward, \& G.C. Sloan, \apjl, (1996), in press.

\bibitem{matt96} K. Matthews, T., Nakajima, S.R. Kulkarni, \& B.R. Oppenheimer, submitted to \apj, 1996.

\bibitem{ew92} E.F. Erickson \& M.W. Werner, Space Science Reviews, 61 (1992) 95.

\bibitem{a94} J.R.P. Angel, Nature, 368 (1994) 203.

\bibitem{mount94} M. Mountain,  R. Kurz, \& J. Oschmann, in {\it The Gemini 8-m Telescope Projects,
S.P.I.E. Proceedings on Advanced Technology Optical Telescopes V}, 2199 (1994) 41.

\bibitem{e92} E.F. Erickson, Space Science Reviews, 61 (1992) 61.

\bibitem{thomp92} R. Thompson, Space Science Reviews, 61 (1992) 69.

\bibitem{ISO94} ISO Observer's Manual Version 2.0, 31 March 1994, prepared by the ISO Science
Operations Team, p6.

\bibitem{leger93} A. Leger, \etal\ , Darwin Mission Concept, proposal to ESA (1993).

\end{thebibliography}
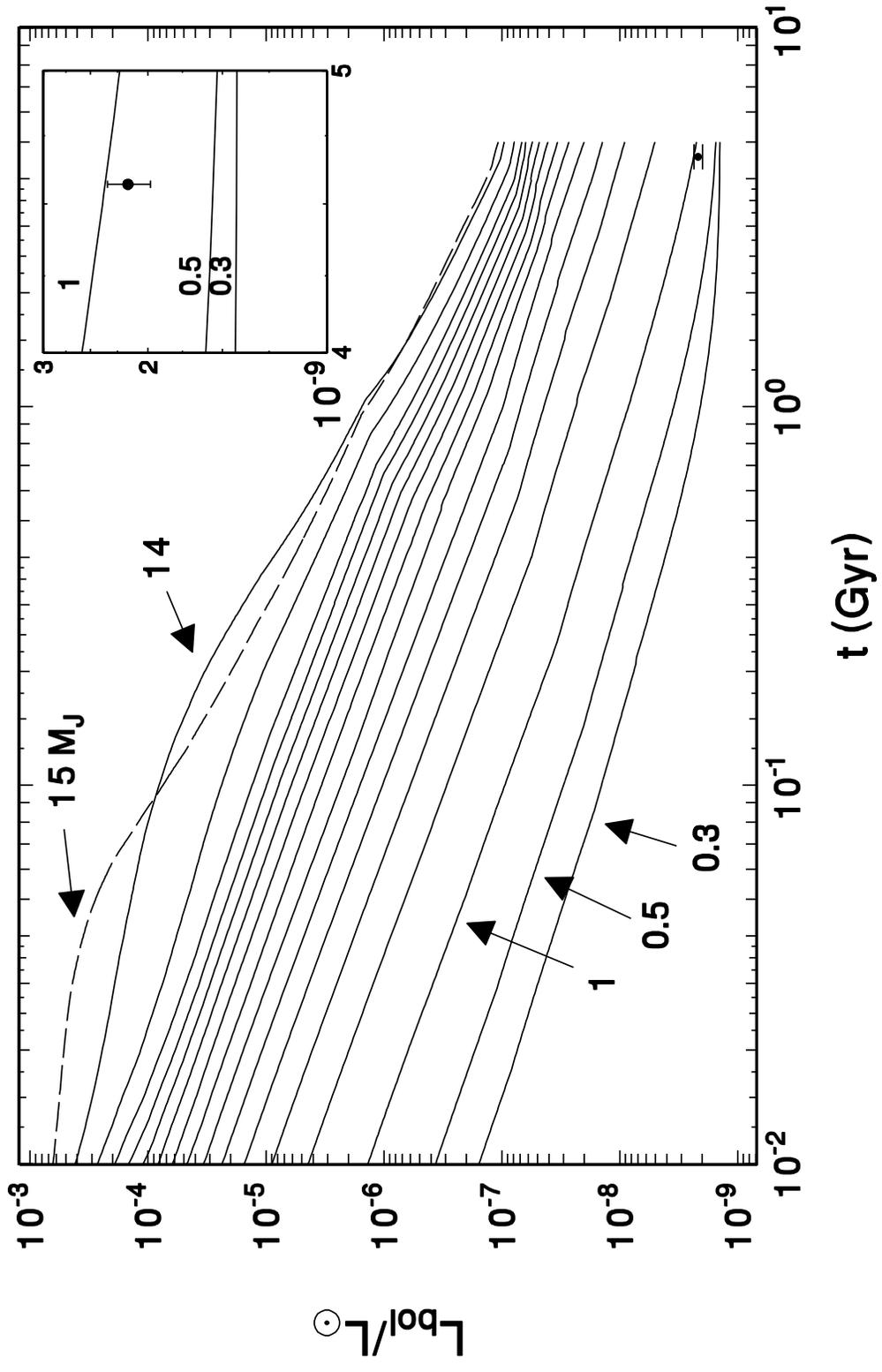
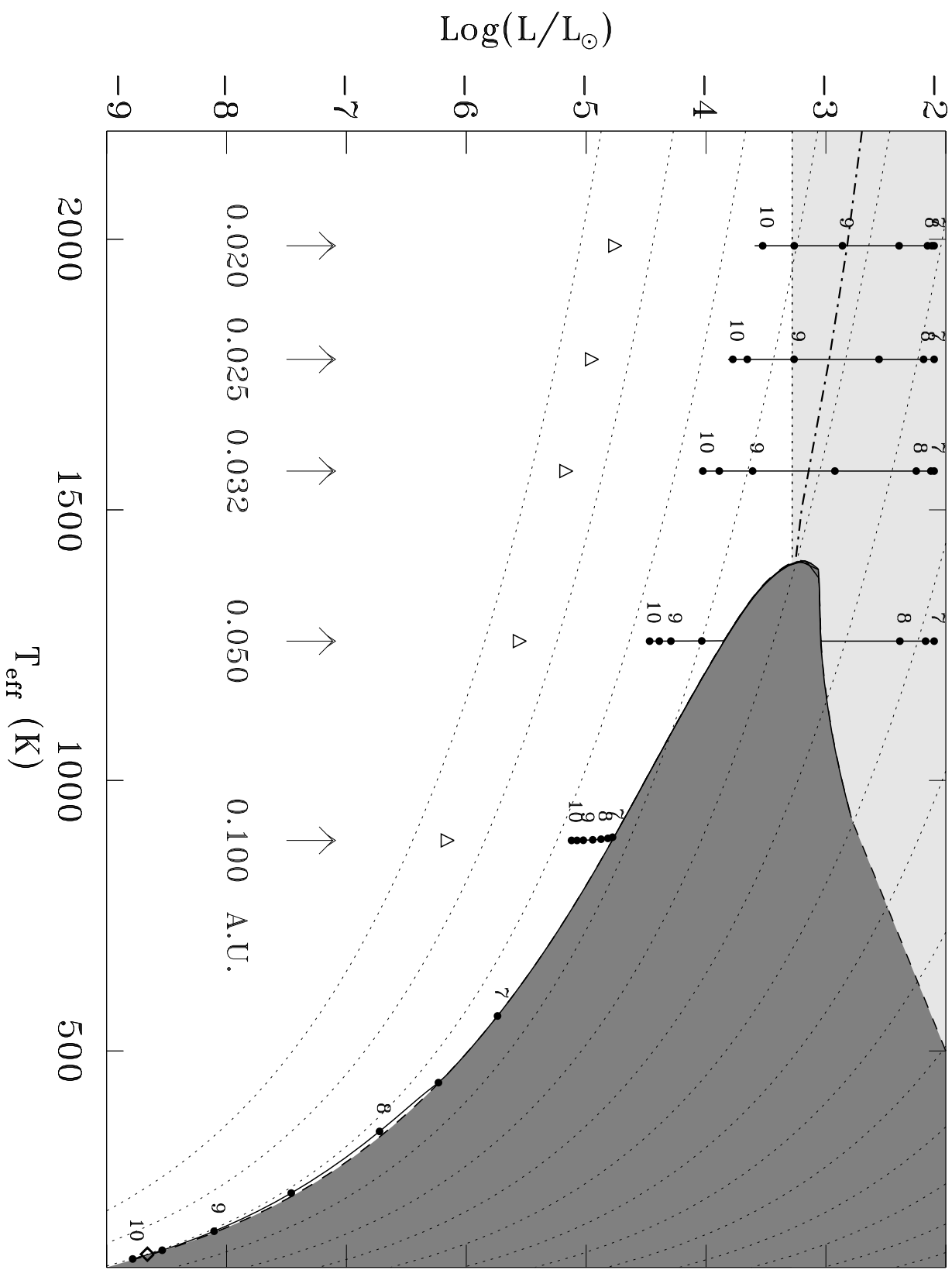
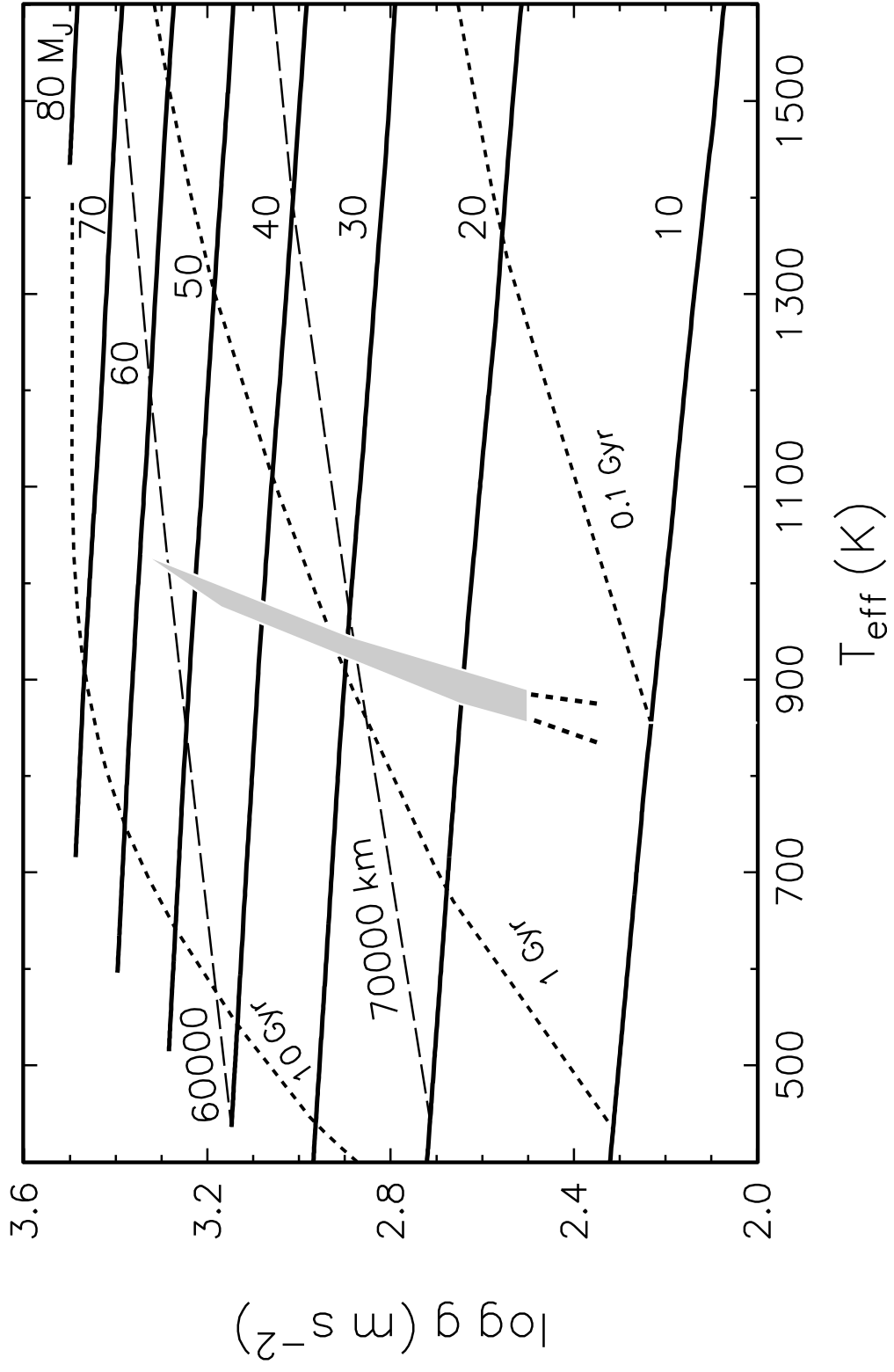
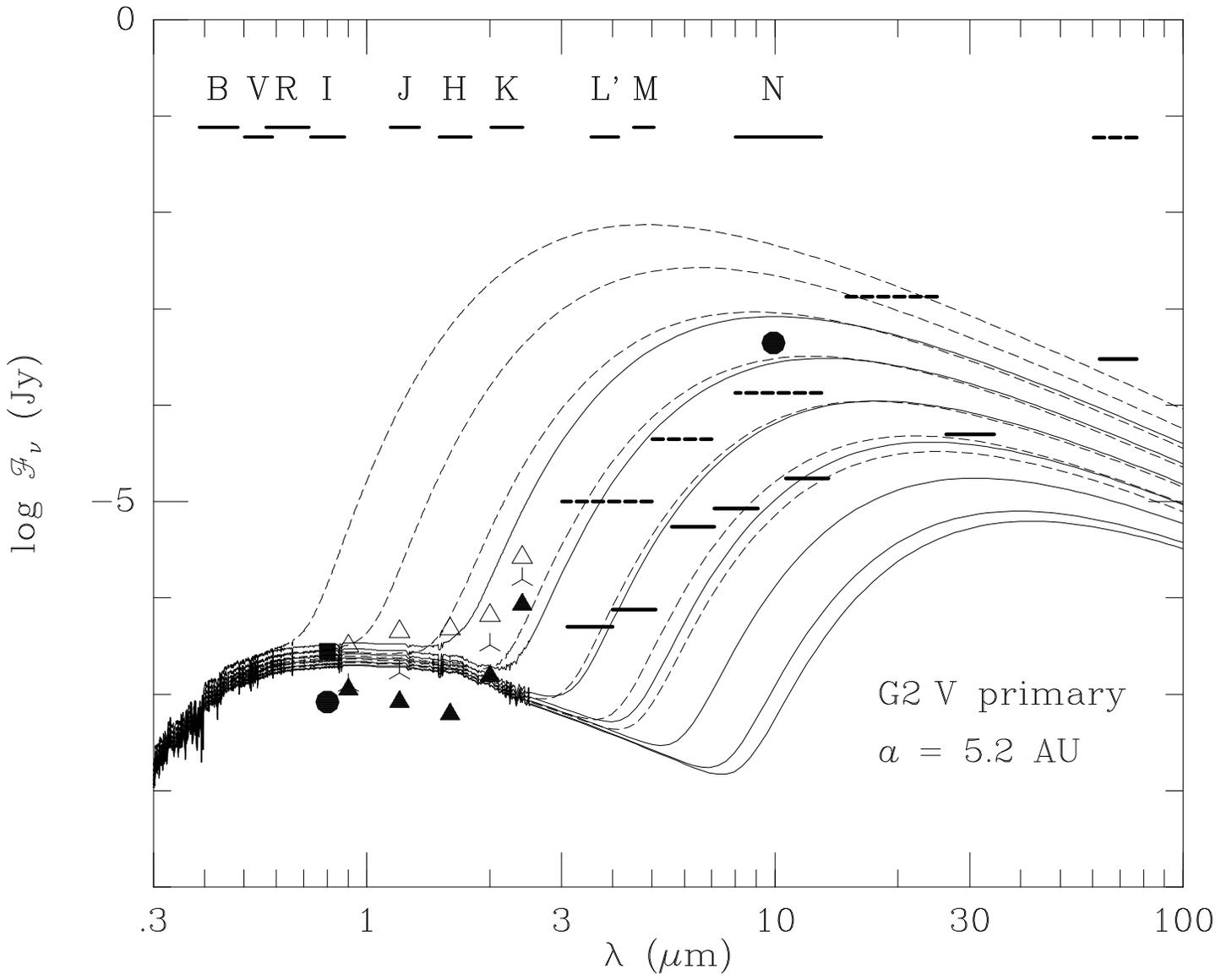
\end{document}